# On the topologic structure of economic complex networks: empirical evidence from large scale payment network of Estonia

Stephanie Rendón de la Torre[*], Jaan Kalda[*], Robert Kitt[*1], Jüri Engelbrecht[*]
[*]Institute of Cybernetics at Tallinn University of Technology, Akadeemia tee 21, 12618, Tallinn, ESTONIA
[1]Swedbank AS, Liivalaia 12, 15038, Tallinn, ESTONIA

______________________________________________________________________________

This paper presents the first topological analysis of the economic structure of an entire country based on payments data obtained from Swedbank. This data set is exclusive in its kind because around 80% of Estonia's bank transactions are done through Swedbank, hence, the economic structure of the country can be reconstructed. Scale-free networks are commonly observed in a wide array of different contexts such as nature and society. In this paper, the nodes are comprised by customers of the bank (legal entities) and the links are established by payments between these nodes. We study the scaling-free and structural properties of this network. We also describe its topology, components and behaviors. We show that this network shares typical structural characteristics known in other complex networks: degree distributions follow a power law, low clustering coefficient and low average shortest path length. We identify the key nodes of the network and perform simulations of resiliency against random and targeted attacks of the nodes with two different approaches. With this, we find that by identifying and studying the links between the nodes is possible to perform vulnerability analysis of the Estonian economy with respect to economic shocks.

*Corresponding author.
E-mail Address: stretomx@gmail.com (S. Rendón de la Torre)

## 1. Introduction

The network approach applied to financial and economic systems has potential to go further on the frontiers of research; there are two currents of origin: one comes from finances, economics and sociology, and the second one comes from computer science, big data challenges, physics, and complex evolving network studies [1]. Both converge in how node representation is done and how the relationships and interactions across the nodes form, whatsoever the nature of these links are. This is an intuitive path that starts to follow the approach that fusions economy and complex systems studies.

Nowadays, networks are a central concept and they can be: biological, technological, economic, social, cultural, among other types. The physical approach has made significant effort during the recent years around the study of evolution and structure of networks [2,3,4,5,6,7,8,9] while some other works have been dedicated to certain network phenomena and specific properties [10,11].

Since the structure of a network has direct influence on the vulnerability and dynamic behavior of the underlying system, important network properties such as stability and robustness can be understood by analyzing the clustering coefficient, the degree distribution and by determining the average shortest path length between nodes in the network [12,13].

In networks, the degree distribution $P(k)$ is the probability that a node links to $k$ number of nodes. Complex networks can be separated into two classes based on their degree distributions:

1) Homogeneous networks are identified by degree distributions that follow an exponential decay. The distribution spikes at an average $k$ and then decays exponentially for large values of $k$, such as the random graph model [14,15] and the small-world model [4], both leading to an homogeneous network: in which each node has approximately the same number of links $k$ and a normal distribution where the majority of the nodes has an average number of connections, and only some or none of the nodes have only some or lots of connections.

2) Heterogeneous large networks or scale-free networks, are those for which $P(k)$ decays as a power law with a characteristic scale. The degree distribution follows a Pareto form of distribution where many nodes have few links and few nodes have many links, therefore, highly connected nodes are statistically significant in scale-free networks.

Network topology gives a fair basis for investigating money flows of customer driven banking transactions. A few recent papers describe the actual topologies observed in different financial systems [13,16,17,18,19]. Other works have focused on shocks and robustness in economic complex networks [20,21,22,23,24].

Scale-free networks display a strong tolerance against random removal of nodes [14] whereas exponential networks not (this means an exponential network can break easily into isolated clusters). Scale-free networks are more resistant to random disconnection of nodes because one can eliminate a considerable number of nodes randomly and the network's structure is preserved and will not break into disconnected clusters. However, the error tolerance is acquired at the expense of survival attack capability. When the most connected nodes are targeted, the diameter of a scale-free network increases and the network breaks into isolated clusters. This occurs because when removing these nodes, the damage disturbs the heart of the system, whereas a random attack is most likely not. One way to entangle the interaction of the nodes is by taking a look to the heavy tail effects they produce and see the implications on their robustness. Heavy-tailed distributions are strong against random perturbations but are extremely sensitive to targeted attacks.

Unlike previous studies, we illustrate the topology of an unstudied complex system that can be analyzed as a particular case of a complex network: Estonia's network of payments. We study the full country economic development, found on Swedbank's data as a proxy. The main goal of our analysis is to study the structure of this economic network. Additionally, this data set is unique given the fact that around 80% of Estonia's bank transactions are done through Swedbank, hence it is expected to reproduce fairly well the structure of the Estonian economy.

This paper is organized as follows. In section 2 we provide the description of the selected data and the methods utilized; section 3 is devoted to the discussion of the results and section 4 concludes the study.

## 2. Materials and Methods
### 2.1 Data

Payment events data from Swedbank AS were used to create the network. Data and information related to identities of the nodes will remain confidential and cannot be disclosed. We believe the utilized data describes fairly well the tendencies of money transactions and is the best possible information available.

The considered dataset corresponds to year 2014. We analyze the network of the payment flows of Swedbank (Estonia), specifically: domestic payments transferred electronically from customer to customer (legal entities). There are 16,613 nodes and 2,617,478 payment transactions in the network. There are 43,375 links if we count them as undirected.

A network (or a graph) is a set of nodes connected by links. The links are the connections between the nodes. In our network, the nodes are the companies and a link is established from one node to another if at least 20 payments were executed, or more than 1,000 money units were paid/received per year. When there is a link from a node to itself, it is called a loop. We eliminated loops resulting from parties making money transfers across their own bank accounts.

There are several ways to define the network of payments; in this study we consider three definitions. The first definition is to look at the structure as a weighted graph where the links have certain weights associated to them representing less or higher important relationships with the nodes. Transactions between any two parties add to the associated link weights in terms of value of payments settled. In this representation we built a payment adjacency matrix that represents the whole image of the network and each element represents the overall money flow traded between companies $i$ and $j$. This non-symmetric matrix represents the weights of the volumes of money exchanged between the companies.

The second definition is to consider an undirected graph, ignoring directions and weights of the payments and considering that two parties are connected if they share at least one payment, then $a_{ij}^u = a_{ji}^u$ and $a_{ij}^u = 1$ if there is a transaction between company $i$ and $j$ or $a_{ij}^u = 0$ if there is no transaction between them. Diagonal elements are equal to 0 and non-diagonal elements are either 0 or 1.

The links can also represent directions on the flow of the relationship. They could be directed or undirected. The third definition is a un-weighted-directed graph where the links follow the flow of money, such that a link is incoming to the receiver and outgoing from the sender of the payment. For this case we have two more matrices, one for the in-degree case and another one for the out-degree case. The choice of the definition of the matrix representation depends on the focus of the analysis.

### 2.2 Components

Depending on how the nodes connect with each other, they can be partitioned into components. A component is a group of nodes such that any two nodes can be connected by a direct or indirect path. A path is a sequence of different nodes, each one connected to the next node. A component of an undirected network is a set of nodes such that for any pair of nodes *i* and *j* there is a path from *j* to *i;* this means that two nodes share the same component if there is a path connecting them. Our analysis treats the network as both undirected and directed, and finds the components and their sizes.

In a directed network the largest component is known as the Giant Weakly Connected Component (GWCC) in which all nodes connect to each other via undirected paths. The core is the Giant Strongly Connected Component in which the nodes can reach each other through a directed path. The Giant Out-Component (GOUT) comprises the nodes that have a path from the GSCC and the Giant In-Component (GIN) comprises the nodes that have a path to the GSCC. The set of disconnected components (DC) are smaller components. Tendrils are nodes that have no directed path to or from the GSCC, but to GOUT and or the GIN [2]. These concepts are shown in Fig. 1.

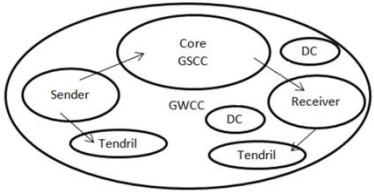

Fig. 1. Components of a directed network.

In order to study the statistical properties and characterize the underlying structure of our network, we use specific useful network metrics [2,3,25].

## 3. Results

Fig. 2(b) displays a picture of the network as a weighted directed graph where each link is shaded by the corresponding weight: with darker shades indicating higher values on the cash flows. The bigger nodes represent those nodes with higher amount of values transferred. Fig. 2(a) shows the network as an undirected representation. This image includes 16,613 nodes and 43,374 links.

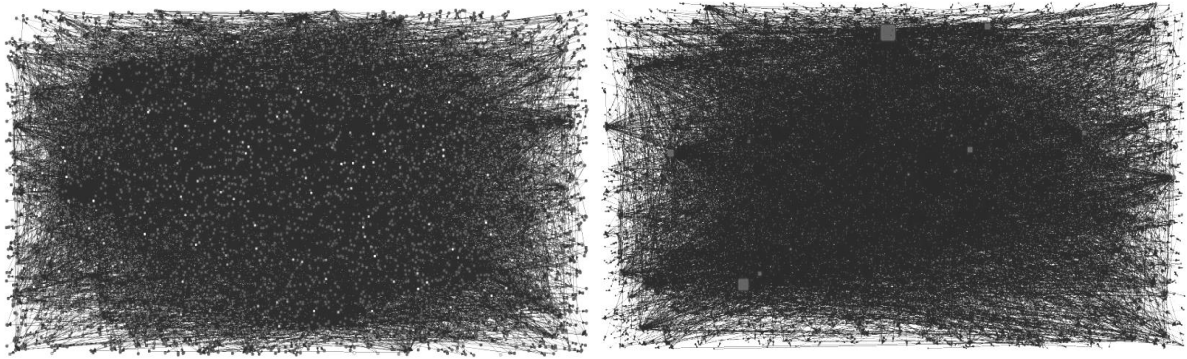

Fig. 2(a) Connectivity network of payments.　　　　　　(b) Weighted-directed representation of the network.

The high number of nodes and links makes difficult to have a detailed visualization of the graph's structure, therefore, we calculate topological and statistical measures that provide a clearer structure of the network.

### 3.1 Topology structure

We find all the components in the undirected graph. We obtained that the GCC is composed by 15,434 nodes which means that 92.8% of the nodes are reachable from one another by following either forward or backward links, suggesting it is a very well connected network. The remaining 7.2% nodes correspond to 508 DC. If we take a directed approach, the GSCC contains 24% of the nodes in the system.

A previous study of the structure of the WWW network components [25] focused on analyzing the robustness of the GCC against removal of nodes, and it was concluded that it is very difficult to destroy the structure of the WWW network by random elimination of links. (Table 3 displays the component sizes of the network of payments, among other statistics).

The degree of a node is defined as

$$k_i = \sum_{j \in \zeta(i)} a_{ij}, \qquad (1)$$

the sum goes over the set $\zeta(i)$ of neighbors of $i$. For example: $\zeta(i) = \{j | a_{ij} = 1\}$.

In a directed network there are two relevant characteristics of a node: the number of links that end at a node and the number of links that start from the node. These quantities are known as the out-degree $k^o$ and the in-degree $k^d$ of a node, and we define them as

$$k^d = \sum_{j \in \zeta(i)} a_{ij}^d, \quad k^o = \sum_{j \in \zeta(i)} a_{ij}^o. \qquad (2)$$

The average degree of a node in a network is the number of links divided by the number of nodes and is defined as

$$\langle k \rangle = \frac{1}{n} \sum k^o = \frac{1}{n} \sum k^d = \frac{m}{n}. \qquad (3)$$

One can categorize networks by the degree distributions shown in the tails. In random networks it is very common to find Poisson distributed links, but in complex system networks it is common to find a distribution that follows a power law

$$P(k_i) \sim k_i^{-\gamma}, k \neq 0, \qquad (4)$$

where $\gamma$ is the scaling exponent of the distribution. This distribution is called scale-free and networks with such a degree distribution are referred to as scale-free networks because have no natural scale and the distribution remains unchanged within a multiplicative factor under a rescaling of the random variables [26].

The average degree of our network is $\langle k \rangle = 20$. Most of the nodes have only 5 or less links, and 45% have only 1 link. Like other real networks, the degree distributions (undirected and directed) of the network of payments follow power laws. Fig. 3 displays the degree distributions. In all the distributions, we found regions that can be fitted by power laws, and this implies that the network has a scale-free structure. (We used the maximum likelihood estimation for obtaining the power law exponents [11]). The degree distribution in Fig. (3a) follows a power law with a scaling exponent:

$$P(\geq k) \propto k^{-2.45}. \qquad (5)$$

The in-degree distribution in Fig. (3b) follows a power law defined as

$$P(k) \sim k^{-2.49}. \tag{6}$$

The out-degree distribution Fig. (3c) follows a power law defined as

$$P(k) \sim k^{-2.39}. \tag{7}$$

In all the cases there is an area at the end of the tail that looks like a cut-off which can be explained by the fact that the system is finite and there is a maximum number of connections that a company could hold.

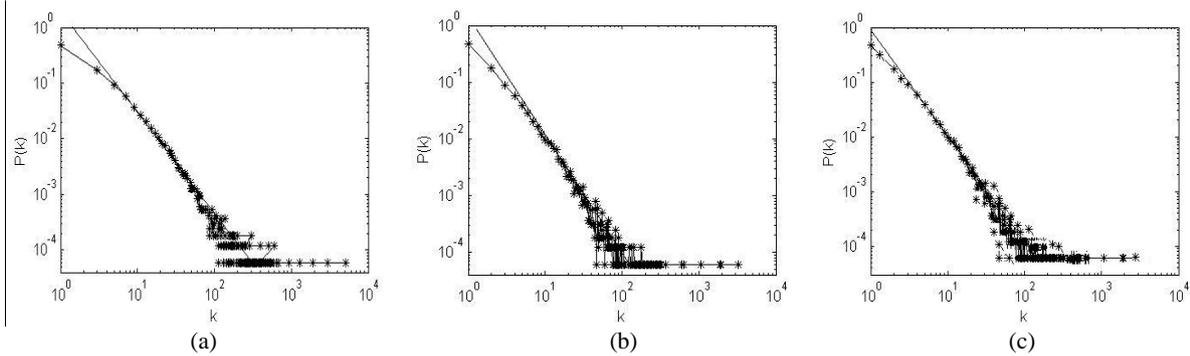

Fig. 3 X axis corresponds to the number *k* of degrees and the Y axis is *P(k)*. (a) Empirical degree distribution for the connectivity network. (b) Empirical in-degree distribution. (c) Empirical out-degree distribution. All the plots are log-log representations of histograms.

In a random network, the degree distributions follow a Poisson distribution. A degree distribution following a power law distribution appears to be a common feature in complex networks such as the World Wide Web, proteins interactions, phone calls and food webs, among others, but also shown in systems of payments of different banks [16,17,18]. The degree distributions obtained here are comparable to those obtained in the aforementioned studies. Table 2 includes a limited list of the power-law exponents obtained in different types of real networks.

**3.2 Weight, strength, size and diameter**

The basic properties of a network are the number of nodes *N* and the overall number of links *k* (Table 1 shows the general characteristics of our network). The number of nodes defines the size of the network while the number of links relative to the number of possible links defines the connectivity of a network.

Connectivity *(p)* is the unconditional probability that two nodes are connected by a direct link. For a directed network, connectivity is defined as

$$(p) = \frac{k}{n(n-1)}. \tag{8}$$

In our case, the connectivity is 0.13 and this means the network is sparse because 87% of the potential connections are not used during the year.

The diameter is the maximum distance between two companies (measured by the number of links) and in our network this distance is equal to 29. This number is substantially higher when compared to the diameter of a random network of comparable characteristics (19) and this big difference in the diameter points to the existence of certain companies that send or receive money to other specific companies, and this contours specific and preferred paths for transactions. Intuitively this makes sense, because for companies in general, it is important to choose carefully their trading partners, clients, service providers or suppliers based on geographical location, affinity in the goals of the companies, cost policies, future joint ventures, agreements or any other reasons.

The strength of the nodes is the sum of the weights of all the links. In this case, the strength measures the overall transaction volume for any given node. The node-weighted strength is defined as

$$s_i = \sum_{j \in \zeta(i)} w_{ij}, \tag{9}$$

where $w_{ij}$ is the weight of the link between nodes $i$ and $j$ and the sum runs over the set $\zeta(i)$ of neighbors of $i$. The average strength can be calculated as a function of the $k$ number of links of a node to examine the bond between the strength and the degree.

Fig. (4a) displays the distribution of link weights weighted by the number of payments transacted. This distribution follows a power law. The same power law relationship occurs between the strength and the degree of a node Fig. (4b) and Fig. (4c). These results were fitted by power laws with the following scaling exponents:

The volume link weight distribution

$$P(w) \sim w^{-1.98}, \tag{10}$$

where the scaling exponent is 1.98.

The volume out-degree strength distribution

$$P(s) \sim s^{-2.21}, \tag{11}$$

where the scaling exponent is 2.21.

The volume out-degree strength distribution

$$P(s) \sim s^{-2.32}, \tag{12}$$

where the scaling exponent is 2.32. There are some deviations from the power law behavior but they are sufficiently small.

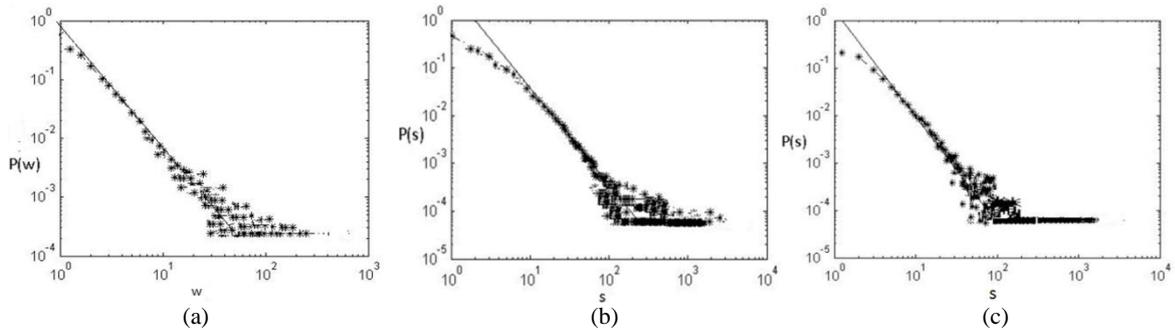

(a) (b) (c)
Fig. 4(a) Link weight distribution by volume. (b) Node in-degree distribution by strength. (c) Node out-degree distribution by strength.

## 3.3 Clustering, betweenness and average shortest path length

The clustering coefficient of a node is the tendency to cluster; is the density around node *i*. It represents the proportion of the closest nodes of a node which are linked to each other.

$$C(i) = \frac{1}{k_i(k_i - 1)} \sum_{j \neq k} a_{ij} a_{jk} a_{ik}. \tag{13}$$

The overall clustering coefficient is the mean of the clustering coefficients $\langle C \rangle$ of all the nodes. It indicates if there is a link between two companies who have a common trading partner. In our case, the average clustering coefficient is 0.183, suggestive of cliquishness in our network. This means that two companies that are trading partners with a third one, have an average probability of 18.3% to be trading partners with one another, than will any two other companies randomly chosen. The clustering coefficient across nodes is highly spread, as seen in Fig. 5.

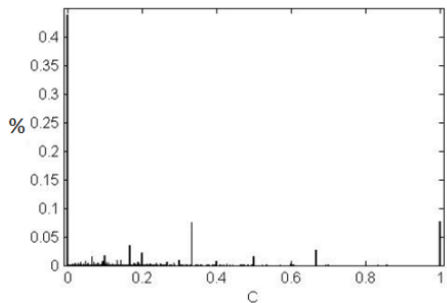

Fig. 5 Distribution of the clustering coefficient.

Fig. 5 shows that more than 52% of the nodes have a clustering coefficient of 0 or 1; therefore, the network is dispersed. There is ~9% probability that two neighbors of a node are linked whereas around 45% are not linked at all. This high level percentage of unlinked neighbor nodes can be explained by the high number of nodes with degrees equal to 1 which are very frequent in scale-free networks.

Compared with other real networks this average clustering coefficient is low (See Table 2 for comparison). In this study, such a coefficient is fair. Business relationships between companies are commonly settled through medium or long term contracts. A company would like to remain doing business with the same parties (suppliers, clients, service providers, institutions, etcetera) because in general, it is easier and cheaper than changing them time after time. A change on a trading party could mean a decrease on profits or an increase on costs. A low clustering coefficient in our payment network reflects this perspective.

In our case, the clustering coefficient is higher than the connectivity, therefore, the network is not random (in a random network the clustering coefficient is equal to the connectivity; a random network is built by randomly adding links to a given set of nodes). A random network of a comparable size has a clustering coefficient around 70 times lower than our network.

Betweenness *σ(m)* of a node *m* is the total number of shortest paths between all possible pairs of nodes that pass through this node; it is a measure of the number of paths between other nodes that run through the node *i*; the more paths this node has, the more central is the node *i* in the network. It indicates whether or not a node is important in the traffic of the network.

$$\sigma(m) \equiv \sum_{i \neq j} \frac{B(i,m,j)}{B(i,j)}, \qquad (14)$$

where $B(i,j)$ is the total number of shortest paths between nodes $i$ and $j$ and the sum goes over all the pairs of nodes for which at least one path exists, with $B(i,j) > 0$. The nodes with high betweenness control the network.

The results in Table 2 show that the average betweenness for the links is 40 and for the nodes is 110, meaning that each company handles in average 110 shortest paths, and the higher is the number of shortest path the more central the company is for the network.

The average shortest path length $<l>$ was calculated with the Dijkstra's algorithm [27]. In the connectivity network this value is equal to 7.1. The network is a small world with 7.1 degrees of separation, so in average any company can be reached by other one only in a few steps. Our network has low connectivity but it is densely connected. This characteristic is in line with the fact that there are companies that act as hubs or key nodes and lead to short distances between the other companies.

93% of the nodes are within 7 links of distance from each other and this suggests that the network of payments is comprised of a core of nodes with whom the other companies interact with. There is a smaller group of 1,081 nodes (6.5% of the total number of nodes in the network) connected by high value links. This group contains weighted links that comprise 75% of the overall value of the funds transferred. Fig. 6 shows the graph of the k-core. A k-core in an undirected graph is a connected maximal induced sub-graph which has minimum degree greater than or equal to $k$. Alternatively, the k-core is the (unique) result of iteratively deleting nodes that have degree less than $k$, in any order.

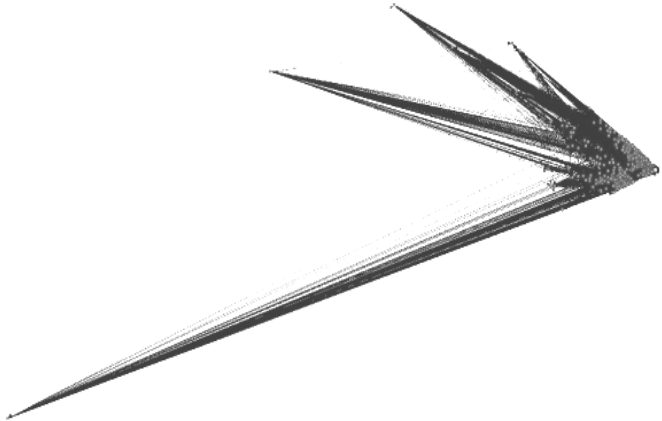

Fig. 6 Graph representation of the k-core.

### 3.4 Robustness simulation and degree correlations

One of the characteristics that makes a hub or a key node an important node is its high betweenness not just its high degree. Hubs often connect groups of clusters of sub-areas of the graph that are not connected to one another directly. These nodes are important because they shorten path lengths making for high reachability and fast movement of information. But they may also be important as brokers and key-players that connect the graph because of their betweenness [28].

In order to gain more understanding on how the network is likely to behave as a whole, let us address the question: if a node were removed, would the structure of the network become divided into disconnected clusters? One can consider several approaches to find the key nodes in the network which may act as enablers among otherwise disconnected groups and we find the nodes that connect the network by locating the vulnerable parts (see Hanneman and Riddle [29]).

A total of 1,401 cut-points or key nodes were found, this means that around 8% of the nodes are relevant to keep the structure connected as it is, or in other terms, if we remove these nodes then the number of components and the average path lengths between the nodes would increase, leaving the network vulnerable to break.

We run a simulation for the GCC that shows random removal of a fraction of nodes and another simulation considering strategically chosen nodes. Then we measure the average shortest path length $<l>$ and the relative size of the GCC as functions of the percentage $d$ of deleted nodes [2,30,31]. The results are displayed in Fig. 7. The effect of the targeted removal of nodes causes a quick growth in the average shortest path length until the GCC disappears, *GCC($p_c$)=0* at a very low level of targeted damage (less than 10%). We will call this level the percolation threshold $p_c$. It is noticeable that a weak but smart attack destroys the network. In the random removal of nodes the damage is less than in the targeted damage. We established that our network of payments has shown scale-free properties, and this kind of networks are resilient to random damage, so it is barely possible to destroy the network of payments by random removal, but if we remove the exact portion of particularly selected nodes, it breaks completely. This effect has been seen in financial systems in economic crisis before: companies or banks may declare in bankruptcy and the whole system stays healthy, but if certain organizations declare in bankruptcy then the whole system collapses.

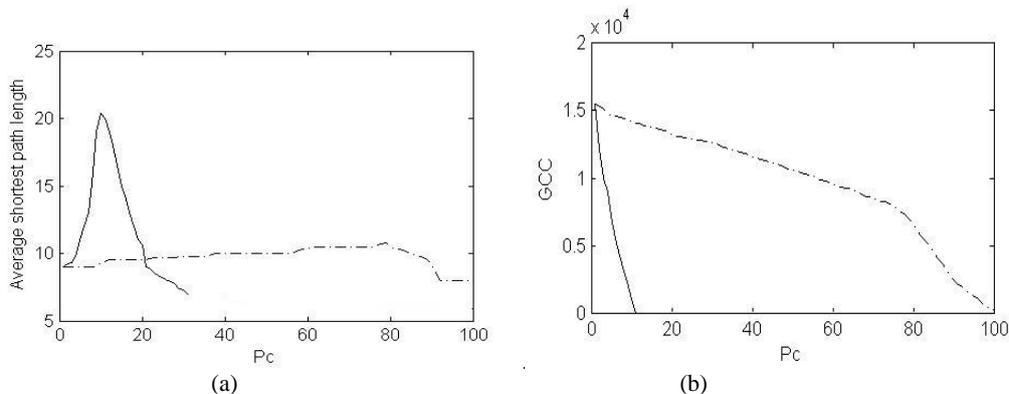

(a) (b)
Fig. 7. Plots of the targeted and random damage over the network of payments. (a) The average shortest-path length $<l>$ in the GCC plotted against the percentage of removed nodes. (b) The GCC plotted against the percentage of removed nodes. Continuous lines display the effect of the targeted removal and the dashed lines display the effect of the random removal of nodes. $p_c$ are the percolation thresholds, for each case.

It is not rare that the GCC in heavy-tailed networks is resilient against random removal of nodes. If the degree distribution of the network is fat-tailed, then this fact determines the topology of the system. However, it might be possible that when removing nodes in a random way, the tail of the degree distribution changes and then the GCC structure would be damaged [2].

There are other heuristic methods available in literature to calculate the optimal percolation threshold of nodes that breaks the network into disconnected clusters (such as high degree node, k-core, closeness and eigenvector centralities). However, a common characteristic in these approaches is that they do not necessarily optimize a measure that reflects the collective influence arising from considering the entire influential nodes at once. Under a collective influence approach, nodes' inherent strength and weakness arises collectively from the configuration of interactions they have with the other components. Morone and Makse [32] designed an algorithm suitable for large networks that has proven to perform better than other empirical methods because it predicts a smaller set of optimal influencer nodes (the nodes that destroy the network if removed).

The collective influence of a node is defined as the product of the node's reduced degree (the number of its nearest connections, $k_i$, minus one), and the total reduced degree of all nodes $k_j$ at a distance $\ell$ from it ($\ell$ is defined as the shortest path)

$$CI_\ell(i) = (k_i - 1) \sum_{j \in \partial \text{Ball}(i,\ell)} (k_j - 1), \qquad (15)$$

where Ball($i, \ell$) is the set of nodes inside a ball of radius $\ell$ around node $i$. $\partial\text{Ball}(i,\ell)$ is the frontier of the ball and comprises the nodes $j$ that are at a distance $\ell$ from $i$. By computing the *CI* for each node, it is possible to find the ones with the highest collective influence and remove them.

The collective influence algorithm maps the problem of optimal influence on the computation of the minimum structural amount of nodes that reduces the largest eigenvalue of the non-backtracking matrix of the network (see Morone and Makse [32]).

We performed a simulation using the *CI* approach, and the results are shown in Fig. 8. We measure the collective influence of a group of nodes as the fall in the size of the GCC which would occur if the nodes in question were eliminated. The figure shows the GCC when a fraction of its nodes is eliminated. The optimal percolation threshold occurs when removing around 6% of the nodes because that is the point where $GCC(p_c)=0$. This result also implies that there are a huge number of companies with a large number of payments which in fact have a minor influence in the whole economic network.

Fig. 8 shows a better performance than previous strategy used on Fig. 7 (which is based on a betweenness centrality and where the optimal percolation threshold is higher than in collective influence method).

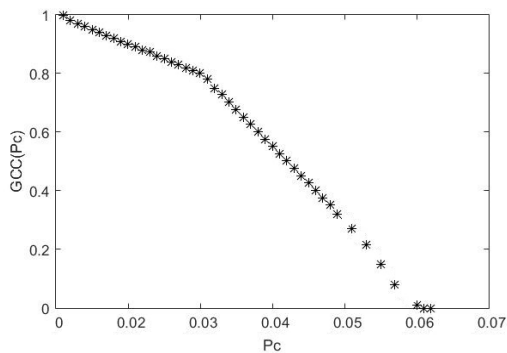

Fig. 8 GCC of the network of payments as a function of the percolation threshold $p_c$ by using the collective influence algorithm.

A practical measure of correlations is the average nearest neighbor degree function. A network is called assortative if its nodes with a certain degree are more likely to have links with nodes of similar degree, and it is called disassortative when the contrary occurs. For example, when low degree nodes are more likely connected with nodes of higher degrees, or when high degree nodes are more likely connected with low degree nodes.

A method that calculates these aforementioned correlation measures is the average nearest neighbor degree function (see Serrano et al. [33,10]). The conditional probability of a node with degree $k$ to be connected to a node of degree $k'$ is defined as

$$P(k'|k) = \frac{P(k',k)}{P(k)}, \tag{16}$$

where $P(k',k)$ is the probability of two nodes, with degree $k'$ and $k$ to be connected by a link. $P(k)$ is the degree distribution. The average nearest neighbor degree function is defined as

$$\langle k_{nn}\rangle(k) = \sum_{k'} k' P(k'|k). \tag{17}$$

Previous studies [2] have shown that social networks usually have significant assortative mixing; biological, technological and other financial networks have shown disassortative mixing [17,24]. Fig. 9 shows the affinity of the connectivity network. The correlation is -0.18, which means there is a negative dependence between the degree of a node and the degrees of its neighbors; therefore the system exhibits disassortative mixing. The function $\langle k_{nn}\rangle(k)$ decreases with $k$ suggesting that the high degree nodes, which are represented by companies who have many business partners such as service providers, clients or suppliers, usually have a large number of links to companies which have only one link (or just few), then the high degree nodes tend to connect with the low degree ones.

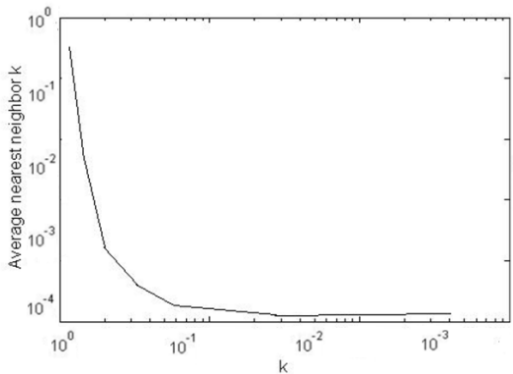

Fig. 9 Affinity of the connectivity network.

Disassortative mixing has implications for network resilience. For example, when this type of mixing is found, the attacks to the highest degree nodes are effective when trying to destroy the network quickly because these nodes are being approximately distributed over the network and forming links on different paths between other nodes, hence, this characteristic makes our network particularly vulnerable to targeted attacks.

## 4. Conclusion

We studied the structure of the economic network of an entire country using Swedbank's payments database. After extracting the network's topology, characteristics and statistics we conclude that this economic network shares many of the features found in empirical complex networks, such as scale-free degree distributions, small world characteristic and low clustering coefficient.

Our results show that this economic network is disassortative in terms of degree. The system shows topological heterogeneity due to its heavy tails and scale-free structure in the degree distributions. This scale-free structure indicates that few companies in Estonia trade with many parties while the majority trade with only few.

In our network, the clustering coefficient is low and disperse (more than 52% of the nodes have either a clustering coefficient of 0 or 1). A low coefficient is a fair result because it shows how companies perceive business partners change as an avoidable expense. A company might prefer to keep working with regular trading partners (for example: service providers, clients or suppliers) for a medium or long term instead of changing them often, in order to save money and time.

The network is a small world with just 7 degrees of separation: in average any company can be reached by other only in a few steps. The connectivity is smaller than the overall clustering coefficient; therefore, the network cannot be classified as random.

Regarding the diameter size of our network: it is high when compared with that of a random network (1.5 times higher). The diameter in our results suggests a preference for specific paths of money flows between companies. This preference refers to companies that trade more with specific parties over others based on decisions relative to costs saving, geographical location, convenience, or any other type of decision.

We performed two separate analyses to reveal the robustness of our economic network. The first one is based on centralities and the second one is based on an approach focused on collective influencer nodes. First, we found the key nodes that prevent the network to break into disconnected components. The simulation for the GCC assuming a targeted removal of key nodes causes a quick growth in the average shortest path length until the GCC disappears at a percolation threshold of 8%, while in the random removal the damage is extremely small. This revealed the robustness of our economic network against random attacks but also revealed its vulnerability to smart attacks.

In the second analysis we followed the collective influence strategy. The percolation threshold is close to 6%; therefore, the performance of the optimal percolation threshold is better when following this method because it reduces the percentage substantially. The interpretation for this low level of optimal percolation threshold is that there are a lot of companies with enormous amounts of payments that have a weak influence in the economic network as a whole, and this also reveals that the most influential companies in the network are not necessarily the most connected ones or those having more economic activity. Both results agree on the fact that a small portion of economic entities maintains the whole network unified.

5. Acknowledgments

Algorithms in UCINET [34] and Pajek [35] were used to calculate network statistics and images. We thank Swedbank AS for enabling us with the data set for the analysis. This research was supported by the European Union through the European Regional Development Fund (project TK 124).

## 6. Tables

Table 1. Network's characteristics

| | |
|---|---|
| **Companies analyzed** | 16,613 |
| **Total number of payments analyzed** | 2,617,478 |
| **Value of transactions** | 3,803,462,026 * |
| **Average value of transaction per customer** | 87,600 * |
| **Max value of a transaction** | 121,533 * |
| **Min value of a transaction (aggregated in whole year)** | 1,000 * |
| **Average volume of transaction per company** | 60 |
| **Max volume of transaction per company** | 24,859 |
| **Min volume of transaction per company (aggregated in whole year)** | 20 |

*All money amounts are expressed in monetary units.

Table 2. Scaling exponents and clustering coefficients for different types of reported networks

| Type | Network | Exponent | Clustering coefficient* | References |
|---|---|---|---|---|
| | **Bank of Japan payments** | $\gamma = 2.1$ | - | [16] |
| | **US Federal Reserve Bank** | $\gamma^i = 2.11$ $\gamma^o = 2.15$ | 0.53 | [17] |
| **Economical** | **Austrian Interbank Market payments** | $\gamma^i = 1.7$ $\gamma^o = 3.1$ | 0.12 | [18] |
| | **WWW** | $\gamma^o = 2.4$ $\gamma^i = 2.1$ | - | [36] |
| | **Peer-to-peer network** | $\gamma = 2.1$ | 0.012 | [37] |
| **Technological** | **Digital electronic circuits** | $\gamma = 3$ | 0.03 | [38] |
| | **Film actors** | $\gamma = 2.3$ | 0.78 | [4] |
| | **Email messages** | $\gamma^i = 1.5$ $\gamma^o = 2.0$ | 0.16 | [39] |
| **Social** | **Telephone call** | $\gamma = 2.1$ | - | [40] |
| | **Protein interactions (yeast)** | $\gamma = 2.4$ | | |
| | **Metabolism reactions** | $\gamma^i = 2.2$ $\gamma^o = 2.2$ | 0.022 0.32 | [41] [42] |
| **Biological** | **Energy lancscape for a 14-atom cluster** | $\gamma = 2.78$ | 0.073 | [43] |

$\gamma^i$ = scaling exponent for in-degree distribution. $\gamma^o$ = scaling exponent for the out-degree distribution. $\gamma$ = scaling exponent for the connectivity distribution. *Refers to average clustering coefficient.

Table 3. Summary of Statistics

| Statistic | Value | *Components* | # of nodes |
|---|---|---|---|
| **N** | 16613 | **GCC** | 15434 |
| **# Payment** | 2617478 | **DC** | 1179 |
| **Undirected Links** | 43375 | **GSCC** | 3987 |
| *< k >* | 20 | **GOUT** | 6054 |
| $\gamma^o$ | 2.39 | **GIN** | 6172 |
| $\gamma^i$ | 2.49 | **Tendrils** | 400 |
| $\gamma$ | 2.45 | **Cutpoints** | 1401 |
| *< C >* | 0.183 | **Bi-component** | 4404 |
| *< l >* | 7.1 | **k-core** | 1081 |
| **C** | 0.13 | | |
| **D** | 29 | | |
| *< σ >* **(nodes)** | 110 | | |
| *< σ >* **(links)** | 40 | | |

N = number of nodes. <k> = average degree. $\gamma^o$ = scaling exponent for the out-degree empirical distribution. $\gamma^i$ = scaling exponent for the in-degree empirical distribution. $\gamma$ = scaling exponent for the connectivity degree distribution. <C> = average clustering coefficient. < l > = average shortest path length. C = connectivity per cent. *D* = Diameter. < σ > = average betweenness.

*References:*

[1] Schweitzer F, Fagiolo G, Sornette D, Vega-Redondo F, White DR. Economic networks: what do we know and what do we need to know? Adv. Comp. Syst. 2009;12: 407-22. Doi:10.1142/S0219525909002337

[2] Dorogovtsev, SN, Mendes, JFF. Evolution of networks. New York: Oxford University Press; 2003.

[3] Albert R, Barabási AL. Statistical mechanics of complex networks. Rev. Mod. Phys. 2002;74:47-97. doi:10.1103/RevModPhys.74.47

[4] Watts DJ, Strogatz, SH. Collective dynamics of small-world networks. Nat. 1998;393:440-442. doi:10.1038/30918

[5] Watts DJ. Six Degrees: The science of a connected age. New York: WW Norton & Company; 2003.

[6] Newman MEJ. Assortative mixing in networks. Phys. Rev. Lett. 2002;89,208701. doi:10.1103/PhysRevLett.89.208701

[7] Barrat A, Barthélemy M, Vespigniani A. Dynamical processes on complex networks. Cambridge, UK: Cambridge University Press; 2008.

[8] Broder A, Kumar R, Maghoul F, Raghavan P, Rajagopalan S, Stata R, et al. Graph structure in the web. Comput. Netw. 2000;33:309-20. doi: 10.1016/S1389-1286(00)00083-9

[9] Faloutsos M, Faloutsos P, Faloutsos C. On power-law relationships of the internet topology. Comput. Commun. Rev. 1999;29:251-62. doi: 10.1145/316194.316229

[10] Catanzaro M, Boguña M, Pastor-Satorras R. Generation of uncorrelated random scale-free networks. Phys. Rev. E 2005;71(2),027103. doi: 10.1103/PhysRevE.71.027103

[11] Newman MEJ. Power laws, Pareto distributions and Zipf's law. Contemp. Phys. 2005;46(5):323-51. doi: 10.1016/j.cities.2012.03.001

[12] Barabási AL, Albert R. Emergence of scaling in random networks. Science 1999;286:509-512. doi: 10.1126/science.286.5439.509

[13] Lublóy A. Topology of the hungarian large-value transfer system. Magyar Nemzti Bank (Central Bank of Hungary) MNB Occasional Papers 2006;57.

[14] Barabási, AL, Albert R, Jeong H. Mean-field theory of scale-free random networks. Physica A 1999;272:173-87. doi:10.1016/s0378-4371(99)00291-S

[15] Erdös P, Rényi A. On random graphs. Publicationes Mathematicae Debrecen 1959; 6:290-7.

[16] Inaoka H, Nimoniya T, Taniguchi K, Shimizu T, Takayasu H. Fractal network derived from banking transactions - An analysis of network structures formed by financial institutions. Bank of Japan Working papers 2004;04-E-04.

[17] Soramäki K, Bech ML, Arnold J, Glass RJ, Beyeler WE. The topology of interbank payment flows. Physica. A 2007;379(1):317-333. doi:10.1016/j.physa.2006.11.09

[18] Boss M, Helsinger H, Summer M, Thurner S. The network topology of the Interbank Market. Quant. Financ. 2004;4(6):677-84.

[19] Rordam KB, Bech ML. The topology of Danish Interbank money flows. Finance Research Unit, 2008, working paper 59.

[20] Deschatres F, Sornette D. The dynamics of book sales: endogenous versus exogenous shocks in complex networks. Phys. Rev. E 2005;72(1) 016112. doi: 10.1103/PhysRevE.72.016112

[21] Iori G, Jafarey S. Criticality in a model of banking crisis. Physica. A 2001;299(1):205-12. doi: 10.1016/S0378-4371(01)00297-7

[22] Crucitti P, Latora V, Marchiori M, Rapisarda A. Error and attack tolerance of complex networks. Physica. A 2005;340(1):388-94. doi: 10.1371/journal.pone.0059613

[23] Kitt R. Economic decision making: application of the theory of complex systems. In: Chaos Theory in Politics / Eds. S. Banerjee, S. S. Ercetin, A. Tekin. Dordrecht: Springer,

2014, Ch. Part II. Politics, Complex Systems, Basin of Attractions, 51-73. doi:10.1007/978-94-017-8691-1_4
[24] Iori G, De Masi G, Precup OV, Gabbi G, Caldarelli G. A network analysis of the Italian overnight money market. J. Econ. Dyn. Control 2007;32(1):259-78. doi:10.1016/j.jedc.2007.01.032
[25] Adamic LA, Huberman BA, Barabási AL, Albert R, Jeong H, Bianconi G. Power-law distribution of the World Wide Web. Science 2000;287;2115a. doi:10.1126/science.287.5461.2115a
[26] Newman MEJ. The structure and function of complex networks. SIAM Rev. 2003; 45(2):167-256. doi: 10.1137/S003614450342480
[27] Gibbons A. Algorithmic graph theory. Cambridge, UK: Cambridge University Press; 1985.
[28] Borgatti SP, Everett MG, Johnson JC. Analyzing social networks. London, UK: SAGE Publications Limited; 2013.
[29] Hanneman RA, Riddle M. Introduction to social network methods. Riverside, CA: University of California Riverside; 2005.
[30] Albert R, Jeong H, Barabási AL. Error and attack tolerance of complex networks. Nat. 1999;406;(6794):378-82. doi :10.1038/35019019
[31] Cohen K, Erez D, Avraham D, Havlin S. Resilience of the internet to random breakdowns. Phys. Rev. Lett. 2000;85:4625-30. doi: 10.1103/PhysRevLett.86.3682
[32] Morone F, Makse HA. Influence maximization in complex networks through optimal percolation. Nat. 2016; 524;65-8. doi: 10.1038/nature14604.
[33] Serrano MA, Maguitman MA, Bogunan A, Fortunato M, Vespignani A. Decoding the structure of the WWW: Facts versus sampling biases. ArXiv:cond-mat, 0511035, 2005.
[34] Borgatti SP, Everett MG, Freeman LC. Ucinet 6 for Windows: software for social network analysis. Harvard, MA: Analytic Technologies; 2002.
[35] Batagelj V, Mrvar A. Exploratory social network analysis with Pajek. Cambridge, UK: Cambridge University Press; 2005.
[36] Albert R, Jeong H Barabási, AL. Diameter of the World Wide Web. Nat. 1999;401:130 1. doi:10.1038/43601.
[37] Ripeanu M, Foster I, Iamnitchi I. Mapping the Gnutella network: properties of large-scale peer-to-peer systems and implications for system design. IEEE Internet Comput. 2002;6(1):50-7.
[38] Ferrer R, RV Cancho, Solé RV. Optimization in complex networks. Statistical mechanics of complex networks, Lect. Notes Phys. 2003(625):114-25. doi: 10.1007/b12331
[39] Ebel H, Mielsch LI, Bornholdt S. Scale-free topology of e-mail networks. Phys. Rev. E 2002;66(3),35103R. Doi: 10.1103/PhysRevE.66.035103
[40] Aiello W, Chung F, Lu L. A random graph model for massive graphs. Proceedings of the Thirty-Second Annual ACM Symposium on Theory of Computing 2000:171-80. doi:10.1145/335305.335326
[41] Jeong H, Mason SP, Barabási AL, Oltvai ZN. Lethality and centrality in protein networks. Nat. 2001;411:41-2. doi:10.1038/35075138
[42] Jeong H, Tombor B, Albert R, Oltvai ZN, Barabási AL. The large-scale organization of metabolic networks, Nat. 2000;407:651-4.
[43] Doye JPK. The network topology of a potential energy landscape: A static scale-free network. Phys. Rev. Lett. 2002;88 238701. doi:10.1103/PhysRevLett.88.238701